\newcommand{\be}{\[}
	\newcommand{\ee}{\]}
	\newcommand{\bea}{\begin{eqnarray*}}	
	\newcommand{\eea}{\end{eqnarray*}}
	\newcommand{\nabbar}{\bar{\nabla}}
	\newcommand{\parhat}{\hat{\partial}_{0}}
	\documentstyle[12pt,leqno]{article}
        \reversemarginpar
\begin{document}
        \begin{center}  
        {\bf Well Posed Reduced Systems for the Einstein Equations\\ }

        \vspace{0.5cm}  
        Yvonne Choquet-Bruhat\\    
        \vspace{0.3cm}
	{\em Gravitation et Cosmologie Relativiste \\
	Universit\'{e} de Paris VI,t.22-12, 75252 \\ }
	\vspace{0.5cm}
	James W. York, Jr.\\
        \vspace{0.3cm}  
        {\em Department of Physics and Astronomy \\    
        University of North Carolina  \\  
	Chapel Hill, NC 27599-3255\\ 
	E-mail: york@physics.unc.edu\\}  
        \vspace{0.3cm}  

        \end{center} 

\rm

\begin{abstract}

We review some well posed formulations of the evolution part of 
the Cauchy problem of General Relativity that we have recently obtained.  
We include also a new first order symmetric hyperbolic system based directly 
on the Riemann tensor and the full Bianchi identities.  It has only physical 
characteristics and matter sources can be included.  It is completely 
equivalent to our other system with these properties.

\end{abstract}

\section{Introduction}

We will review some recently obtained well posed formulations of the 
evolution part of the Cauchy problem (see \cite{kn:CBY1}) in General 
Relativity, considered as the time history of the two fundamental 
forms of the geometry of a spacelike hypersurface, its metric $\bar{g}$ 
and its extrinsic curvature $K$.  On such an hypersurface, for instance 
an ``initial'' one, these two quadratic forms must satisfy four equations, 
called constraints.  The constraints can be posed and solved as an 
elliptic system by known methods. 

The proof of the existence of a causal evolution in local Sobolev 
spaces of $\bar{g}$ and $K$ into an Einsteinian spacetime does not however 
result directly from the equations giving the time derivatives 
of $\bar{g}$  and K in terms of space derivatives of these quantities 
in a straightforward 3+1 decomposition of the Ricci tensor of the 
spacetime metric, which contains also the lapse and shift characterizing 
the time lines.  These equations do not appear as a hyperbolic system for 
arbitrary lapse and shift, in spite of the fact that their characteristics 
are only the light cone and the time axis (see \cite{kn:HF}). 

In this paper, we review different methods used recently for obtaining a 
hyperbolic evolution system for these geometrical unknowns which manifest 
their propagation governed by the light cone.  Particularly interesting 
are first order symmtric hyperbolic systems, since they seem to be the 
most amenable to numerical computations.  We construct explicitly such 
a system, using a harmonic time slicing condition.  Its characteristics 
are the physical light cone and the time direction orthogonal to the space 
sub-manifolds of the chosen slicing.  Among the propagated quantities is, 
in effect, the Riemann curvature.  The space coordinates and the shift are 
arbitrary and, in this sense, the system is gauge invariant.  In fact, we 
can make the lapse also take arbitrary values by introducing a given 
arbitrary function in its definition.  

\section{3+1 Decomposition of the Riemann and Ricci Tensors}

We suppose in all this paper that the spacetime is a smooth 
manifold, $V=M \times \Re$, endowed with a metric $g$ of signature 
$(-,+,+,+)$, such that the ``slices'' are spacelike.  This hypothesis 
is no restriction for globally hyperbolic spacetimes.

We choose on $V$ a moving coframe such that the dual frame has a time 
axis orthogonal to the slices $M_{t}$ while the space axes are tangent 
to them, namely we set
\begin{eqnarray*}
\theta^{0} = dt  \\
\theta^{i} = dx^{i} + \beta^{i} dt
\end{eqnarray*}
with $t \in \Re$ and $x^{i}$, $i=1,2,3$ local coordinates on $M$.  The 
Pfaff derivatives $\partial_{\alpha}$ with respect to $\theta^{\alpha}$ are 
\begin{eqnarray*}
\partial_{0} \equiv \frac{\partial}{\partial t} - \beta^{i} \partial_{i}  \\
\partial_{i} \equiv \frac{\partial}{\partial x^{i}}   .
\end{eqnarray*} 
In this coframe, the metric $g$ reads 
\be
ds^{2} = g_{\alpha \beta} \theta^{\alpha} \theta^{\beta} \equiv - 
N^{2}(\theta^{0})^{2} + g_{ij}\theta^{i}\theta^{j}   .
\ee
The $t$-dependent scalar $N$ and space vector $\beta$ are called the 
lapse and shift of the slicing.  Any spacetime tensor decomposes into 
sets of time dependent space tensors by projections on the tangent space 
or the normal to $M_{t}$.

We define for any $t$-dependent space tensor $T$ another such tensor, 
$\parhat T$, of the same type by setting 
\be
\parhat \equiv \frac{\partial}{\partial t} - {\cal L}_{\beta}
\ee
where ${\cal L}_{\beta}$ is the Lie derivative on $M_{t}$ with respect to 
$\beta$.

The extrinsic curvature of the slices is the time dependent space tensor 
given by 
\be
K_{ij} \equiv -\frac{1}{2} N^{-1} \parhat g_{ij}
\ee
We denote by an overbar geometric quantities associated with the Riemannian 
metric $\bar{g} \equiv g_{ij}dx^{i}dx^{j}$ induced by $g$ on the slices.  

The Riemann curvature tensor of $g$ admits the following decomposition 
into time dependent space tensors (cf. \cite{kn:Lich}, zero shift; 
\cite{kn:CB}, arbitrary shift; \cite {kn:Y1}, spacetime viewpoint)
\bea
R_{i\:\:kl}^{\:\:j} = \bar{R}_{i\:\:kl}^{\:\:j}-
K^{j}_{k}K_{li}+K^{j}_{l}K_{ki}  \\ 
R_{i\:\:kl}^{\:\:0} = N^{-1}(\nabbar_{k}K_{li} - \nabbar_{l}K_{ki})   \\
R_{i \:\:  j0}^{\:\: 0} = -(N^{-1}\parhat K_{ij} + K_{im}K^{m}_{j} + 
N^{-1}\nabbar_{j}\partial_{i}N) 
\eea
From these formulas result the following ones for the Ricci curvature, 
where we have denoted by $H$ the mean extrinsic curvature of the space 
slices, $H = K^{i}_{i}$:

\bea
R_{ij} & \equiv & -N^{-1} \parhat K_{ij} + HK_{ij} - 2K_{im}K^{m}_{j} - 
N^{-1}\bar{\nabla}_{j}\partial_{i}N + \bar{R}_{ij}   \\
R_{0j} & \equiv & N(-\bar{\nabla}_{h}K^{h}_{j} + \partial_{j}H)   \\
R_{00} & \equiv & N(\bar{\nabla}^{i}\partial_{i}N - NK_{ij}K^{ij} + 
\partial_{0}H)
\eea 

From these identities results the following one for the Einstein tensor, 
leading to the so-called "Hamiltonian constraint": 

\be
G_{00} \equiv R_{00}-\frac{1}{2}g_{00}R \equiv 
\frac{1}{2}N^{2}(\bar{R}-K.K+H^{2}) 
\ee

\section{Wave Equations for K. Local Existence Theorems}

\newcounter{test}

Using the relation between $\parhat g_{ij}$ and $K_{ij}$, we obtain 
(with no factor $\frac{1}{2}$ in the index symmetrization)
\bea
\parhat \bar{R}_{ij}   & \equiv &   -\nabbar^{h} \nabbar_{(i}(NK_{j)h}) 
+ \nabbar_{h} \nabbar^{h}(NK_{ij}) + \nabbar_{j} \partial_{i}(NH)    \\
	             & \equiv &   -N \nabbar_{(i} \nabbar^{h} K_{j)h} - 
\nabbar^{h} \partial_{(i} NK_{j)h} - 2 N \bar{R}^{h}_{\:\:ijm}K^{m}_{h} 
- N \bar{R}_{m(i} K^{m}_{j)} +  \\
                     &        & \nabbar_{h} \nabbar^{h} (NK_{ij}) + 
\nabbar_{j} \partial_{i}(NH)   . 
\eea 
We now use the expression for $R_{0i}$ and $R_{ij}$ to obtain the identity
\bea
\Omega_{ij} & \equiv & \parhat R_{ij} - \nabbar_{(i}R_{j)0} \\
            & \equiv & -\parhat(N^{-1}\parhat K_{ij}) + \parhat(HK_{ij} - 
2K_{im}K^{m}_{j}) - \parhat(N^{-1} \nabbar_{j} \partial_{j}N) - \\
	    &        &  N \nabbar_{i} \partial_{j} H - 
\nabbar_{(i}(K_{j)h} \partial^{h} N) - 
2 N \bar{R}^{h}_{\:\:ijm} K^{m}_{h} - N \bar{R}_{m(i} K^{m}_{j)} + \\
            &        & \nabbar_{h} \nabbar^{h} (N K_{ij}) + 
H \nabbar_{j} \partial_{i} N
\eea
This identity shows that for a solution of the Einstein equations

\vspace{0.25cm}
$(E) \hspace{4cm} R_{\alpha \beta} = \rho_{\alpha \beta}$
\vspace{0.25cm}

\noindent the extrinsic curvature $K$ satisfies a second order 
differential system which is quasi diagonal with the principal part 
the wave operator, except for the terms $\nabbar_{i} \partial_{j} H$.  
The other unknowns $\bar{g}$ and $N$ appear at second order except for 
the term $\parhat \nabbar_{j} \partial_{i} N$.  These facts have been 
used in two different ways to obtain for $K$ a quasi diagonal, wave 
type system. 

\indent 1.  One eliminates at the same time the third derivatives of $N$ 
and the second derivatives of $H$ (cf. in the case of zero 
shift \cite{kn:CBR} and for arbitrary shift \cite{kn:CBY2}) by 
requiring $N$ to satisfy 

\vspace{0.25cm}
$(N')  \hspace{3.9cm}   \partial_{0} N + N^{2} H = 0   $
\vspace{0.25cm}

\noindent The second order equation for $K$ reads then as a wave type 
nonlinear system: 

\vspace{0.25cm}
$(K')  \hspace{3.9cm} N \Box K_{ij} = N Q_{ij} + \Theta_{ij} $
\vspace{0.25cm}

\noindent where we set 
\bea 
\Box K_{ij} & \equiv & -N^{-2} \parhat \parhat K_{ij} + 
\nabbar^{h} \nabbar_{h} K_{ij}  \\
NQ_{ij}     & \equiv & -K_{ij} \partial_{0} H + 
2 g^{hm} K_{m(i} \parhat K_{j)h} + 
4 N g^{hl} g^{mk} K_{lk} K_{im} K_{jh} + \\
            &        & (2\nabbar_{(i} K_{j)l}) \partial^{l} N - 
2 H N^{-1} \partial_{i} N \partial_{j} N - 
2 \partial_{(i}N \partial_{j)}H - \\
            &        & 3 \partial_{h} N \nabbar^{h} K_{ij} - 
K_{ij} \nabbar^{h} \nabbar_{h} N - 
N^{-1} K_{ij} \partial^{h} N \partial_{h} N + \\
            &        & N^{-1} K_{h(i} \partial_{j)} N \partial^{h} N + 
(\nabbar_{(i} \partial^{h} N) K_{j)h} + 
2 N \bar{R}^{h}_{\:\:ijm} K^{m}_{h} + \\
            &        & N \bar{R}_{m(i} K^{m}_{j)} - 
2 H \nabbar_{j} \partial_{i} N   \\
\Theta_{ij} & \equiv &  \parhat \rho _{ij} - \nabbar_{(i} \rho_{j)0} 
\eea 
The equation $(N')$ expresses that the time coordinate is harmonic. 

Using the expression for $H$ we see that $(N')$ reads 
\be
\parhat log(N/(det \bar{g})^{\frac{1}{2}}) = 0 
\ee
The general solution of this equation is 
\be 
N = \alpha^{-1}(det \bar{g}^{\frac{1}{2}})
\ee
where $\alpha$ is an arbitrary tensor density such that 
\be
\parhat \alpha = 0   . 
\ee
The algebraic expression of the harmonic time-slicing condition is 
called ``algebraic gauge.''

If we replace $N$ by the value obtained above in the second order 
equation for $K$ we obtain a quasi diagonal system with principal 
part the wave operator, with terms depending on $\bar{g}$ and its 
derivatives of order $\leq 2$.  This system reduces to a third order 
hyperbolic system when we replace $K$ by $-(2N)^{-1} \parhat \bar{g}$.  
The local existence theorem of Leray for the solution of hyperbolic 
systems gives immediately the local in time existence of solutions of 
this reduced system, in local in space Sobolev spaces, with the smallest 
index known for generic solutions of Einstein's equations, and domain of 
dependence determined by the light cone.  It can be proved 
(see \cite{kn:AACBY2}) that a solution of the reduced system 
on $M \times I$ is a solution, in algebraic gauge, of the full 
Einstein equations if the initial data satisfy the Einstein 
equations on $M_{0}$.

\vspace{0.5cm}
Remark 1.  If we take for $\alpha$ an arbitrary given function we still 
obtain a quasi diagonal system for $K$, with additional a priori given 
terms.   Such a generalization is equivalent to replacing $(N')$ by 
the equation

\vspace{0.25cm}
$(N'')    \hspace{3.7cm}   N^{-1} \partial_{0} N + N H = f $
\vspace{0.25cm}

\noindent with $f$ an arbitrary function. 

2.  Replace in the term $\nabbar_{j} \partial_{i} H$ the mean 
curvature $H$ by an a priori given function $h$ (a procedure used 
in \cite{kn:CK} with $h=0$, in the asymptotically euclidean case).  
The equations $\Omega_{ij} = \Theta_{ij}$ become then, when $N$ is 
known, a quasi diagonal second order system  for $K$ with principal 
part the wave operator, namely: 

\vspace{0.25cm}
$(K'') \hspace{3.7cm} \Box K_{ij} = P_{ij} + \Theta_{ij}  $
\vspace{0.25cm}

\noindent where 
\be 
\Box K_{ij} \equiv - \parhat \parhat (N^{-1} K_{ij}) + 
\nabbar^{h} \nabbar_{h} (N K_{ij})
\ee
Here, $P_{ij}$ depends only on $K$ and its first derivatives, 
on $\bar{g}$, $N$ and $\partial_{0} N$ together with their space 
derivatives of order $\leq 2$.  It is given by 
\bea 
P_{ij} & \equiv & \parhat (-H K_{ij} + 2 g^{hm} K_{(im} K^{m}_{j)}) 
+ \parhat (N^{-1} \nabbar_{j} \partial_{i} N) + 
\nabbar^{h}(\partial_{(i} N K_{j)h}) + \\
       &        & 2 N \bar{R}^{h}_{\:\:ijm} K^{m}_{h} +  
N \bar{R}_{m(i} K^{m}_{j)} - H \nabbar_{j} \partial_{i} N + 
N \nabbar_{j} \partial_{i} h, 
\eea
while $\Theta_{ij}$, zero in vacuum, is 

\be
\Theta_{ij} \equiv \parhat \rho_{ij} - \nabbar_{(i} \rho_{j)0}  .
\ee

When $\beta$, $N$ and the sources $\rho$ are known the above 
equation together with 

\vspace{0.25cm}
$(g')  \hspace{3.7cm} \parhat g_{ij} = -2 N K_{ij} $
\vspace{0.25cm}

\noindent are again a third order quasi diagonal system 
for $\bar{g}$, hyperbolic if $N > 0$ and $\bar{g}$ is properly Riemannian. 

The condition $H = h$ is a ``mean curvature'' gauge choice 
which imposes, through the equation $R^{0}_{0} = \rho^{0}_{0}$, 
that $N$ satisfy the following elliptic equation on each slice 

\vspace{0.25cm}
$(N) \hspace{3.0cm} \nabbar^{i} \partial_{i} N - (K_{ij} K^{ij} 
- \rho^{0}_{0}) N = - \partial_{0} h  $
\vspace{0.25cm}

Note that for energy sources satisfying the strong energy condition 
we have $-\rho^{0}_{0} \geq 0$, as well as 
$K.K \equiv K_{ij} K^{ij} \geq 0$, an important property for 
the solution of the elliptic equation.  
In the given mean curvature gauge the unknowns $N$ and $\bar{g}$ 
satisfy, for any choice of the shift $\beta$, a mixed elliptic and 
hyperbolic system. 

Local existence theorems (i.e. in a neighborhood of $M$ 
in $M \times \Re$) have been proven 
(see \cite{kn:CBY2}, \cite{kn:CBY3}) when $M$ is 
compact and when $M$ is asymptotically euclidean. 

\section{Case of Compact M}

Theorem 1.   {\em Let $(M,e)$ be a smooth compact 
Riemannian manifold.  Let there be given on 
$M \times I$, with $I \equiv [0,T]$, a ``pure space'' 
smooth vector field $\beta$ and a function $h$ such that
\be
h \in \bigcap_{2 \leq k \leq 3} C^{3-k}(I,H_{k}),\:\:\:\:\: 
\partial_{0}h \geq 0, \:\:\:\:\: \partial_{0}h \not\equiv 0  .
\ee
There exists an interval $J \equiv [0,\ell]$, $\ell \leq T$ such 
that the system $(K'')$, $(g')$, (N) has one and only one solution 
on $M \times J$
\bea
\bar{g} & \in & \bigcap_{1 \leq k \leq 3} C^{3-k}(J,H_{k}) \\
K       & \in & \bigcap_{1 \leq k \leq 2} C^{2-k}(J,H_{k}) \\
N       & \in & \bigcap_{0 \leq k \leq 2} C^{2-k}(J,H_{2+k})   
\eea
with $N > 0$ and $\bar{g}$ uniformly equivalent to $e$, taking the 
initial data
\bea
g_{ij}(0,.) & = & \gamma_{ij} \in H_{3} \\
K_{ij}(0,.) & = & k_{ij} \in H_{2}
\eea
if $\gamma$ is a properly riemannian metric uniformly equivalent to 
$e$ and $k.k \not\equiv 0$.
}
\section{Case of (M,e) Euclidean at Infinity}

The manifold $M$ is the union of a compact set and a finite number of 
disjoint sets (its ``ends'') diffeomorphic to the exterior of a ball 
in $\Re^{3}$; the smooth given metric $e$ reduces on each end to the 
euclidean metric.  The weighted Sobolev space $H_{s,\delta}$ on $(M,e)$ 
is the completion of $C^{\infty}_{0}$ in the norm
\be
||f||_{H_{s,\delta}} \equiv ( \int_{M} \Sigma_{ _{0 \leq k \leq s}} 
\sigma^{2k+2\delta} | D^{k}f |^{2} \mu(e))^{\frac{1}{2}}
\ee
where $\sigma^{2} \equiv 1+d^{2}$ and $d$ is the distance in the metric 
$e$ to some fixed point in $M$. 

The use of the theory of elliptic equations on an asymptotically euclidean 
manifold in $H_{s,\delta}$ spaces (see \cite{kn:CBC}) and of weighted 
energy estimates leads to the following theorem: 

\vspace{0.5cm}
Theorem 2. {\em  The system (1),(2),(3) with Cauchy data $\gamma$ and 
$k$ on the manifold $(M,e)$ euclidean at infinity with given $\beta$ 
and $h$ on $M \times [0,T]$ has one and only one solution $(\bar{g},K,N)$ 
on $M \times J$, $J \equiv [0,\ell]$ a sufficiently small subinterval 
of $I$, if the Cauchy data are such that $\gamma -e \in H_{3,-1}$ and 
is uniformly equivalent to $e$ while $k \in H_{2,0}$ and
 $\dot{k} \in H_{1,1}$.  The solution belongs on each $M_{t}$ to 
the same functional spaces as the data.  It is such that $N > 0$ 
and $\bar{g}$ is uniformly equivalent to $\gamma$. 
}
\vspace{0.5cm}

Remark 2.  The estimates show in fact that $\bar{g} - \gamma$ is 
such that for each $t$ we have: 
\be
\bar{g}_{t} - \gamma \in H_{3,0}
\ee
which is a stronger asymptotic fall off than 
$\gamma - e$ or $\bar{g}_{t} - e$: this property is 
related to the Arnowitt-Deser-Misner theorem of mass conservation.

\section{First Order System (Vacuum)}

We use the wave equation satisfied by $K$ to show 
(see \cite{kn:CBY2}, \cite{kn:AACBY1}) that in spacetime 
{\em dimension 4} a solution of the vacuum Einstein equations,
 together with the harmonic time gauge condition, satisfies a first 
order symmetric system, hyperbolic if $\bar{g}$ is properly 
Riemannian and $N > 0$.  Such a system could be useful to establish 
a priori estimates relevant to global problems.  It may be important 
for numerical computations (see \cite{kn:BMSS}) because symmetric 
hyperbolic systems occur in many areas of mathematical physics, in
 particular in fluid dynamics, and effective codes have been developed 
to study such systems. 

We have obtained for the unknowns $\bar{g}$, $K$, $N$ the equations 

\vspace{0.25cm}
$(1) \hspace{3.0cm}   \parhat g_{ij} = -2 N K_{ij}  $
\vspace{0.25cm}

$(2)  \hspace{3.0cm}   \partial_{0} N = -N^{2} H  $
\vspace{0.25cm}

$(3)  \hspace{3.0cm}   \Box K_{ij} \equiv Q_{ij} $
\vspace{0.25cm}

\noindent To obtain a first order system we take as additional unknowns: 
\bea 
N^{-1} \parhat K_{ij} = L_{ij}   \\
\nabbar_{h} K_{ij} = M_{hij}     \\
\partial_{i} logN = a_{i}        \\
N^{-1} \parhat \partial_{i} log N = a_{0i}   \\
N^{-1} \nabbar_{j} \partial_{i}log N = a_{ji}
\eea
We take as equation $(3')$ 

\vspace{0.25cm}
$(3') \hspace{3.0cm}    \parhat K_{ij} = N L_{ij} $
\vspace{0.25cm}

\noindent The equation $(3)$ gives 

\vspace{0.25cm}
$(4)  \hspace{3.0cm}   \parhat L_{ij} - N \nabbar^{h} M_{hij} = 
N (H L_{ij} - Q_{ij})  $
\vspace{0.25cm}

In three space dimensions the Riemann tensor is a linear function of 
the Ricci tensor: 
\be
\bar{R}_{lijm} \equiv g_{lj} \bar{R}_{im} + g_{im} \bar{R}_{jl} - 
g_{ij} \bar{R}_{lm} - g_{lm} \bar{R}_{ij} - \frac{1}{2}(g_{lj} g_{im} - 
g_{ij} g_{lm}) \bar{R}
\ee
Using the equation $R_{ij} = 0$ to express $\bar{R}_{ij}$, we write $Q_{ij}$ 
as a polynomial in the unknowns and $g^{ij}$. 

By using the identity 
\be
\parhat \bar{\Gamma}^{h}_{ij} \equiv \nabbar^{h} (NK_{ij}) -
 \nabbar_{(i}(NK^{h}_{j)})
\ee
we see that for an arbitrary covariant vector $u_{i}$ we have 
\be 
\parhat \nabbar_{h} u_{i} \equiv \nabbar_{h} \parhat u_{i} + 
u_{l} (\nabbar_{h}(N K^{l}_{i}) - \nabbar_{i}(NK^{l}_{h}) - 
\nabbar^{l}(NK_{ih}))
\ee
From an analogous formula for tensors we deduce the equation 

\vspace{0.25cm}
$ (5) \hspace{1cm} \parhat M_{hij} - N \nabbar_{h} L_{ij}  =  
N[ a_{h} L_{ij} + (M_{h(i}^{\:\:\:\:l} K_{j)l} + 
K_{l(i} M_{j)h}^{\:\:\:\:l}  - K_{l(j} M_{\:\:i)h}^{l}) + $

$     \hspace{3cm}    K_{l(j}(K^{l}_{i)}a_{h} + a_{i)}K^{l}_{h} - 
a^{l}K_{i)h})  $
\vspace{0.25cm}

\noindent On the other hand $(2)$ implies 

\vspace{0.25cm}
$(6) \hspace{3.0cm} \parhat a_{i} = N a_{0i} = -N ( M_{ih}^{\:\:\:\:h} + 
H a_{i}) $
\vspace{0.25cm}

\noindent while $a_{hi}$ and $a_{0i}$ must satisfy 

\vspace{0.25cm}
$(7) \hspace{0.5cm} \parhat a_{hi} - N \nabbar_{h} a_{0i} = 
N(a_{l}(M_{hi}^{\:\:\:\:l} + M_{ih}^{\:\:\:\:l} - M^{l}_{\:\:ih}) + 
a_{l}(K^{l}_{i} a_{h} + K^{l}_{h} a_{i} - a^{l} K_{ih}) + a_{h} a_{0i}) $
\vspace{0.25cm}

The condition $(2)$ together with the Einstein equation $R^{0}_{0}$ 
imply that $N$ satisfies the inhomogeneous wave equation (this step 
brings in the Hamiltonian constraint determined by $G_{00}$) 
\be
\partial_{0} \partial_{0} N - N^{2} \nabbar^{h} \nabbar_{h} N = 
-N^{3} K_{ij} K^{ij} + 2 N^{3} H^{2} 
\ee
Hence by differentiation, use of the Ricci formula, the definitions of 
$a_{hi}$ and $a_{0i}$ and simplification through the use of $R^{0}_{0} = 0$ 

\vspace{0.25cm}
$(8) \hspace{0.5cm} \parhat a_{0i} - N \nabbar^{h} a_{hi}   =     
N(2a^{h}a_{ih} - 2 K_{hl}M_{i}^{\:\:hl} + H M_{ih}^{\:\:\:\:h} -
 \bar{R}^{h}_{i} a_{h}) + $

$ \hspace{3cm} Na_{i}(H^{2} + 2 a^{h}_{h} + 2 a_{h}a^{h} - 2 K_{hj} K^{hj})$
\vspace{0.25cm}

We use again the equation $R_{ij} = 0$ to replace $\bar{R}_{ij}$ 
by its value in terms of the unknowns.  We have obtained a first 
order system (1,2,$3'$,4,5,6,7,8) in all the unknowns.  The right 
hand sides are polynomial in the unknowns and $g^{ij}$; they do not 
depend on their derivatives.  The left hand sides are linear operators 
on all the unknowns.  Their coefficients depend on these unknowns, and 
not on their derivatives except for the derivatives of $\bar{g}$ which 
appear through the Christoffel symbols $\bar{\Gamma}^{h}_{ij}$.  We can 
use the identity given above to write their evolution equation 
\be
\parhat \bar{\Gamma}^{h}_{ij} = N(M^{h}_{\:\:\:ij} - M_{(ij)}^{\:\:\:\:h} 
- a^{h} K_{ij} - a_{(i}K^{h}_{j)}) . 
\ee
However, if we wish to display an explicitly covariant quasi linear first 
order system for all the unknowns, in particular if $M$ is not diffeomorphic 
to $\Re^{3}$, we introduce on $M$ an a priori given connection $E$, possibly 
derived from a metric $e$ which will also be used to define the Sobolev 
spaces on $M$.  The derivatives in the connection of $\bar{g}$ and the 
given connection differ by terms linear in a tensor $S$, the difference 
of the two connections.  The evolution of this tensor is given by 
\be
\parhat S^{h}_{ij} = N(M^{h}_{\:\:ij} - M_{(ij)}^{\:\:\:\:\:h} - 
a^{h} K_{ij} - a_{(i}K^{h}_{j)}) - \parhat E^{h}_{ij}   . 
\ee
We have now obtained a quasi linear first order system covariant for 
all the unknowns.  Its characteristic matrix, obtained by replacing in 
the principal matrix the operator $\partial$ by a covariant vector $\xi$,
 consists of blocks around the diagonal, some reduced to one element 
$\xi_{0}$ and some $4 \times 4$ matrices with determinant
 $\xi_{0}^{2} - N^{2} \xi^{i} \xi_{i}$.  The characteristics 
are the light cone and the axis orthogonal to the time slices. 
 On the other hand, the system can be symmetrized by multiplication by
 a matrix consisting of blocks around the diagonal equal to one element,
 $1$, or the matrix $(g^{ij})$.  In other words, the system is a first
 order {\em symmetrizable hyperbolic system, with domain of 
dependence determined by the light cone}.  Known local existence
 theorems apply to such a system. 

\section{Bianchi Equations}

Instead of taking as new unknowns the first derivatives of $K$ to
 obtain a first order symmetric hyperbolic system, one can use the
 Riemann tensor of the spacetime metric, which is linear in these 
derivatives, to rewrite the system in Section 6.  It satisfies the 
Bianchi identities: 
\be
\nabla_{\alpha} R_{\beta \gamma , \lambda \mu} +
 \nabla_{\beta} R_{\gamma \alpha , \lambda \mu} +
 \nabla_{\gamma} R_{\alpha \beta , \lambda \mu} \equiv 0.   
\ee
These identities imply by contraction and use of the symmetries 
of the Riemann tensor 
\be
\nabla_{\alpha} R^{\alpha}_{\:\:\mu, \beta \gamma} +
 \nabla_{\gamma} R_{\beta \mu} + \nabla_{\beta} R_{\gamma \mu} \equiv 0 .
\ee
If the Ricci tensor $R_{\alpha \beta}$ satisfies the Einstein equations
\be
R_{\alpha \beta} = \rho_{\alpha \beta}
\ee
then the previous identities imply the equations 
\be
\nabla_{\alpha} R^{\alpha}_{\:\:\mu,\beta \gamma} =
 \nabla_{\beta} \rho_{\gamma \mu} - \nabla_{\gamma} \rho_{\beta \mu}
\ee

The first equations with $(\alpha \beta \gamma) = (ijk)$ and the 
last one with $\mu = 0$ do not contain derivatives of the Riemann 
tensor transversal to $M_{t}$; they are considered as constraints. 

In a particular case (see \cite{kn:Y2}) of the general system considered 
below, the conservation of an initially vanishing Riemann tensor in vacuum
 was obtained.  That result does not require analyticity when it is
 written in symmetric hyperbolic form as below.

We wish first to show that the remaining equations are, for $n=3$ in
 the vacuum case, when $g$ is given, a symmetric first order hyperbolic
 system for the double two-form $R_{\alpha \beta , \lambda \mu}$.
  For this purpose, following Bel (see \cite{kn:BEL}) we introduce
 two pairs of ``electric'' and ``magnetic'' space tensors associated 
with a spacetime double two-form $A$, 
\bea
N^{2} E_{ij} \equiv A_{0i,0j}  \\
D_{ij} \equiv \frac{1}{4} \eta_{ihk} \eta_{jlm} A^{hk,lm}  \\
N H_{ij} \equiv \frac{1}{2} \eta_{ihk} A^{hk}_{\:\:\:\:\:  ,0j}  \\
N B_{ji} \equiv \frac{1}{2} \eta_{ihk} A_{0j,}^{\:\:\:\:\:   hk}
\eea
where $\eta_{ijk}$ is the volume form of $\bar{g}$.  It results from 
the symmetry of  the Riemann  tensor $R$ with respect  to its first 
and second pairs of indices ($R$ is a ``symmetric double two-form'')
 that if $A \equiv R$, then $E$ and $D$ are symmetric while
 $H_{ij} = B_{ji}$.  The Lanczos identity (see \cite{kn:LAN}) for a
 symmetric double two form like $R$, with a tilde representing the 
spacetime double dual, is 
\be
\tilde{R}_{\alpha \beta, \lambda \mu} +
 R_{\alpha \beta, \lambda \mu} = C_{\alpha \lambda} g_{\beta \mu} -
 C_{\alpha \mu} g_{\beta \lambda} + C_{\beta \mu} g_{\alpha \lambda} -
 C_{\beta \lambda} g_{\alpha \mu}
\ee
where $C_{\alpha \beta} = R_{\alpha \beta} - \frac{1}{4}g_{\alpha \beta}R$.
  It follows that when $R_{\alpha \beta} = \lambda g_{\alpha \beta}$, 
then $E = -D$ and $H=B$.  In order to avoid introducing unphysical 
characteristics, and to be able to extend the treatment to the non 
vacuum case, we do not use these properties in the evolution equations,
 but write them as a first order system for an arbitrary double 
2-form $A$ as follows
\bea
\nabla_{0} A_{hk,0j} + \nabla_{k} A_{0h,0j} - \nabla_{h} A_{0k,0j} = 0 \\
\nabla _{0} A^{0}_{\:\:\:i,0j} + \nabla_{h} A^{h}_{\:\:\:i,0j} =
 \nabla_{0} \rho_{ji} - \nabla \rho_{0i}
\eea
and analogous equations with the pair (0j) replaced by (lm).  One 
obtains a first order system for the unknowns $E, H, D, B$ by using
 the following relations deduced from the definition of these tensors 
\bea
A_{hk,0j} \equiv N \eta^{i}_{\:\:hk} H_{ij} \\
A_{hk,lm} \equiv \eta^{i}_{\:\:hk} \eta^{j}_{\:\:lm} D_{ij} \\
A_{0j,hk} = N \eta^{i}_{\:\:hk} B_{ji}
\eea
The system obtained has a principal matrix consisting of 6 identical 
6 by 6 blocks around the diagonal, which are symmetrizable and hyperbolic.
  Hence, the system is symmetric  hyperbolic, when $g$ is a given metric
 such that $\bar{g}$ is properly Riemannian  and $ N > 0$. 

To relate the Riemann tensor to the metric $g$ we use the definition
\be
\parhat g_{ij} = -2 N K_{ij}
\ee

We consider two possibilities: each introduces a choice of gauge for N. 

1.  We fix the mean curvature of the space slices, i.e. we set 
\be
H = h
\ee
with $h$ being a given function.  We deduce from the identity 
giving $R^{\:0}_{i\:\: kl}$ the ``elliptic'' identity 
\be
\nabbar^{h} \nabbar_{h} K_{ij} - \bar{R}^{h}_{j} K_{hi} - 
\bar{R}_{hijm} K^{hm} \equiv \nabbar^{h}(N R^{0}_{\:\:i,jh}) - 
\nabbar_{j}(N^{-1}R_{0i}) + \nabbar_{i} \partial_{j} H  . 
\ee
We write an elliptic system linking the symmetric tensor $K$ with a 
double 2-form $A$ when $H = h$ by symmetrization of the above identity, 
replacement of $R$ by $A$, $H$ by $h$, and the spacetime Ricci tensor 
by $\rho$, zero in vacuum.  Namely, we consider  the system 
\be
\nabbar^{h} \nabbar_{h} K_{ij} - \bar{R}^{h}_{(i}K_{j)h} -
 \bar{R}_{hijm} K^{hm} = \nabbar^{h}(N A^{0}_{(i,j)h}) -
 \nabbar_{(i}(N^{-1}\rho_{j)o}) + \nabbar_{i} \partial_{j} h  . 
\ee
This system is to be solved globally on each slice $M_{t}$, like 
the equation for $N$ in mean curvature gauge.  We have obtained a 
mixed elliptic-hyperbolic system for the unknowns 
$\bar{g}, N, K, E, D, H, B$.  It can be proved that this system 
is well posed in the case of compact or asymptotically euclidean 
$M$, as long as the elliptic operators are injective.  This property 
holds for metrics $\bar{g}$ in a neighborhood of a metric $\gamma$ for
 which it holds, for example if $M \equiv \Re^{3}$ and $\gamma$ is the
 euclidean metric, or if $M \equiv S^{3}$ and $\gamma$ is the canonical
 metric of $S^3$.  

2.  We choose the algebraic form of the harmonic time-slicing condition, 
that is we set $N \equiv \alpha^{-1}|det \bar{g}|^{1/2}$, with $\alpha$ a
 given tensor density.  We follow the ideas used by Friedrich 
(see \cite{kn:HF}) for the Weyl tensor to write a symmetric hyperbolic
 system for $K$ and $\bar{\Gamma}$, namely we consider the following 
identities, deduced from the definition of $K$ and the $3+1$ decomposition 
of the Riemann tensor: 
\be
\parhat \bar{\Gamma}^{h}_{ij} + N \nabbar^{h} K_{ij} = 
K_{ij} \partial^{h}N - K^{h}_{(i}\partial_{j)}N - 
R^{\:\:\:\:\:\:\:\:\:h}_{0(i,j)}
\ee
and 
\be
\parhat K_{ij} + N \bar{R}_{ij} + 
\nabbar_{j} \partial_{i} N \equiv -2 N R^{0}_{\:\:i,0j} - N H K_{ij} + 
N R_{ij}. 
\ee
We obtain equations relating $\bar{\Gamma}$ and $K$, for a given 
double 2-form A, by replacing in these identities the Riemann tensor 
by $A$ and the Ricci tensor of spacetime by a given tensor $\rho$, 
zero in vacuum.  To deduce from this system a symmetric hyperbolic 
first order system,  with the algebraic form of the harmonic gauge,
 one uses the fact that in this gauge one has
\be
\bar{\Gamma}^{h}_{ih} = \partial_{i} log N + \partial_{i}log\alpha 
\ee
The second set of identities leads then to the following equations: 
\bea
\parhat K_{ij} + N \partial_{h} \bar{\Gamma}^{h}_{ij} &  =   &  
N(\bar{\Gamma}^{m}_{ih}\bar{\Gamma}^{h}_{jm} - (\bar{\Gamma}^{h}_{ih} - 
\partial_{i} log \alpha)(\bar{\Gamma}^{l}_{jl} - \partial_{j} log \alpha))+ \\
                                                      &      &N \nabbar_{j}
 \partial_{i} log \alpha -  2 N A^{0}_{\:\:i,0j} - N H K_{ij} + N \rho_{ij}
\eea
The system obtained for $K$ and $\bar{\Gamma}$ has a characteristic matrix 
composed of 6 blocks around the diagonal, each block a 4 by 4 matrix that 
is symmetrizable hyperbolic, with characteristic polynomial 
$\xi^{2}_{0} \xi^{\alpha} \xi_{\alpha}$. 

The whole system for $A, K, \bar{\Gamma}, \bar{g}$ is symmetrizable
 hyperbolic, with characteristics the light cone and the normal to $M_{t}$. 

Remark.  It is somewhat involved to prove that a solution of the 
constructed system satisfies the Einstein equations if the initial data,
 $\gamma$, $k$, satisfy the constraints, but we can argue as follows.
  We consider the vacuum case with initial data $\gamma$ and $k$ 
satisfying the Einstein constraints.  These initial data determine the
 initial values of $\bar{\Gamma}$, and also, if $\beta$ and $N$ are 
known, the initial values of $A_{ij,hm}, A_{jh,i0}, A_{i0,jh}$ by using
 the decomposition formulas.  (We set $A$ equal to the Riemann tensor
 on the initial surface.)  We use the Lanczos formula to determine
 $A_{i0,j0}$ initially.  We know that our symmetrizable hyperbolic
 system has one and only one solution.  Since a solution of Einstein's
 equations in algebraic gauge, proved to exist in a preceding section,
 satisfies together with its Riemann tensor the present system and takes
 the same initial values, that solution coincides with the solution of the
 present system in their common domain of existence.  

\section{A Non-Strict Hyperbolic System with Arbitrary Lapse and Shift}

Lemma. {\em  The following combination of derivatives of components of the
 Ricci tensor of an arbitrary spacetime is a quasi diagonal hyperbolic
 operator for the extrinsic curvature of the space slices: 
\be
\Lambda_{ij} \equiv  \parhat \parhat R_{ij} - \parhat \nabbar_{(i}R_{j)0} +
 \nabbar_{j} \partial_{i} R_{00}
\ee
}

Proof.  A straightforward computation  using previous results shows that 
\bea
\Lambda_{ij} &  \equiv   & \parhat \Box K_{ij} +
 \parhat \parhat(HK_{ij}-2K_{im}K^{m}_{j}) -
 \parhat \parhat(N^{-1}\nabbar_{j} \partial_{i}N) + \\
             &           & \parhat(-\nabbar_{(i}(K_{j)h} \partial^{h} N) -
 2 N \bar{R}^{h}_{\:\:ijm}K^{m}_{h}-N\bar{R}_{m(i}K^{m}_{j)} +
 H\nabbar_{j} \partial_{i}N)+ \\
             &           &  \nabbar_{j} \partial_{i}(N\bar{\Delta}N - 
N^{2}K.K) + {\cal C}_{ij}
\eea
with 
\be
\Box K_{ij} \equiv -\parhat(N^{-1} \parhat K_{ij}) +
 \nabbar^{h} \nabbar_{h}(NK_{ij}), \hspace{2cm} \bar{\Delta} =
 \nabbar_{h} \nabbar^{h}
\ee
and 
\be
{\cal{C}}_{ij} \equiv \nabbar_{j} \partial_{i}(N\partial_{0} H) -
 \parhat(N\nabbar_{j}\partial_{i}H)
\ee

We see that ${\cal C}_{ij}$ contains terms of at most second order in
 $K$ (and also in $N$) and first order in $\bar{g}$ (replace
 $\parhat g_{ij}$ by $-2NK_{ij}$).

The identity given above shows that for a solution of the Einstein
 equations the extrinsic curvature $K$ satisfies, for
 {\em any choice of lapse and shift}, a third order differential system
 which is quasi diagonal with principal part the hyperbolic operator
 $\parhat \Box$.  The other unknown $\bar{g}$ appears at second order
 except for terms appearing through $\nabbar_{j} \partial_{i} \bar{\Delta}N$. 

The system for $\bar{g}$ and $K$ is not hyperbolic in the usual sense
 of Leray because of the third order space derivatives of $\bar{g}$
 appearing in $\nabbar_{j} \partial_{i} \bar{\Delta}N$.

\vspace{0.5cm}
Theorem 3. {\em  The system for $\bar{g}$, $K$ is equivalent to a 
system {\em hyperbolic non strict} in the sense of Leray-Ohya with 
local existence of solutions in Gevrey classes and domain of dependence 
determined by the light cone.}

\vspace{0.5cm}
Proof.  Replace in the equation $K$ by $-(2N)^{-1} \parhat \bar{g}$: this
 gives a quasi diagonal system for $\bar{g}$, but with multiplicity 2 for
 $\partial_{0}$ in the principal operator. 
\vspace{0.5cm}

The system for $\bar{g}$, $K$ can be turned into a {\em hyperbolic system} 
by a {\em gauge choice} (or redefinition of $N$) leading either to an 
elliptic or to a hyperbolic equation for $N$ containing an arbitrary 
function.  Here, we sketch a forthcoming more detailed treatment 
(see \cite{kn:AACBY3}). 

1.  We give arbitrarily a constant $c$, taken to be zero in the 
asymptotically flat case and strictly positive if $M$ is compact.  
 We choose arbitrarily a smooth function $f$ on $M \times \Re$, with
 appropriate asymptotic behavior if $M$ is euclidean at infinity 
and $f \leq 0$ and $f \not \equiv 0$ on each $M_{t}$ if they are 
compact.  We require $N$ to satisfy on each $M_{t}$ the elliptic 
equation depending on $\bar{g}$ 
\be
\bar{\Delta} N - c N = f
\ee
and we determine $N_{0} = \partial_{0}N$ and 
$N_{00} = \partial^{2}_{00}N$ from the elliptic equations obtained
 by differentiating the equation for $N$ and replacing $\parhat \bar{g}$
 by $-2NK$.  It can be shown that the {\em mixed hyperbolic and elliptic}
 system obtained for $\bar{g}, K, N, N_{0}, N_{00}$ is well posed, in
 appropriate functional spaces.  

2.  We choose arbitrarily a smooth function  $f$ on $M \times \Re$ and
 require $N$ to satisfy the wave equation.
\be
N^{-2}\partial_{0}\partial_{0}N - \bar{\Delta}N = f
\ee
We use this wave equation to reduce the term 
$\partial^{2}_{00}(N^{-1} \nabbar_{j}\partial_{i}N)$ in $\Lambda_{ij}$.  
It can be shown that the system for $\bar{g}, K, N, N_{0}$ 
is {\em hyperbolic}, hence well posed in local Sobolev spaces, 
if $N_{0}$ is determined by the hyperbolic equation obtained by 
differentiating the wave equation satisfied by $N$.

\end{document}